%% file: main.tex
\documentclass{article}
\usepackage{spconf,amsmath,amssymb,graphicx,hyperref}
\usepackage{microtype}

\title{CircleMatch: Prototype Matching with Circular Temporal Statistics for Tiny Keyword Spotting}

\name{Jiajun Sun, Zhe Gao\sthanks{Corresponding author: Zhe Gao.}}

\address{Shanghai Normal University, Shanghai, China}

\usepackage{booktabs}
\begin{document}
	\ninept
	\maketitle

\input{sections/01_abstract}
	\input{sections/02_introduction}
	\input{sections/03_method}
	\input{sections/04_experiments}
	\input{sections/05_conclusion}

	% Below is an example of how to insert images. Delete the ``\vspace'' line,
	% uncomment the preceding line ``\centerline...'' and replace ``imageX.ps''
	% with a suitable PostScript file name.
	% -------------------------------------------------------------------------

	% To start a new column (but not a new page) and help balance the last-page
	% column length use \vfill\pagebreak.
	% -------------------------------------------------------------------------
	%\vfill
	%\pagebreak

	\vfill\pagebreak

	% References should be produced using the bibtex program from suitable
	% BiBTeX files (here: strings, refs, manuals). The IEEEbib.bst bibliography
	% style file from IEEE produces unsorted bibliography list.
	% -------------------------------------------------------------------------
	% Enable these two lines after adding the first citation in the text.
	%\bibliographystyle{IEEEbib}
	%\bibliography{refs}
	
	\bibliographystyle{IEEEbib}
	\bibliography{refs}
	
\end{document}

%% file: sections/01_abstract.tex
\begin{abstract}
	Keyword spotting (KWS), the task of identifying predefined
	words in speech, is a core capability of voice-enabled devices.
	Achieving high KWS accuracy under tight parameter budgets
	across different vocabulary sizes remains challenging.
	We present CircleMatch, a matching framework enabling
	KWS with very few parameters.
	Its encoder independently compresses frequency bands
	and fuses them into frame features.
	These features are then matched against learned class-specific
	prototypes to produce temporal response curves.
	Parameter-free circular aggregation encodes time as angles
	and summarizes response distributions and relative timing
	for classification.
	We develop four tiny variants, Circle-D4, Circle-D8,
	Circle-D16, and Circle-D32, ranging from approximately
	1k to 7k parameters in the 12-class setting.
	Experiments with multiple random seeds on Speech Commands
	v1/v2 and the English and Spanish Micro subsets of the
	Multilingual Spoken Words Corpus demonstrate competitive
	accuracy with tiny models.
	Our qualitative analysis further suggests approximate
	shift equivariance of prototype responses and adaptation
	to temporal compression.
	Code and model weights are available at
	\url{https://github.com/ora942878/CircleMatch}.
\end{abstract}


%% file: sections/02_introduction.tex
\section{Introduction}
\label{sec:intro}

Keyword spotting (KWS) identifies predefined words in speech
and is commonly deployed on resource-constrained edge devices.
Such applications require high recognition accuracy while
keeping the number of model parameters as small as possible.
We therefore investigate how to achieve accurate recognition
under extremely limited parameter budgets.

The Google Speech Commands (GSC) dataset~\cite{warden2018speech}
is a widely used spoken-word benchmark for KWS.
Many compact KWS architectures have been developed and
evaluated on this dataset.
Depthwise-separable CNNs~\cite{zhang2017hello} target
microcontroller-based recognition.
MatchboxNet~\cite{majumdar2020matchboxnet} similarly separates
temporal and channel processing.
Residual CNNs~\cite{tang2018residual} use skip connections
to support deeper acoustic models.
Combining factorization with residual learning,
DS-ResNet~\cite{xu2020depthwise} also incorporates channel attention.
For efficient temporal modeling,
TC-ResNet~\cite{choi2019temporal} uses one-dimensional convolutions.
BC-ResNet~\cite{kim2021broadcasted} integrates frequency
information through broadcasted residual connections.
TinySpeech~\cite{wong2020tinyspeech} explores compact
representations using attention condensers.
SparkNet~\cite{svirsky2024sparse} learns sparse representations
for a linear classifier.
These methods seek to preserve discriminative acoustic
information through efficient convolutional designs or
compact learned representations, balancing recognition
accuracy against model size and computational cost.

In audio processing, frequency bands provide a useful way
to organize and characterize acoustic information.
Sub-band CNNs~\cite{kao2019subband} process different frequency
regions with separate filters and fuse their outputs for KWS.
This suggests exploiting frequency-band structure when
designing compact acoustic encoders.

Related KWS research also explores matching and temporal
aggregation.
Deep template matching~\cite{zhang2020template} compares
speech with keyword templates for configurable KWS.
State-sequence pooling~\cite{lopatka2020state} incorporates
sequential structure into KWS training.
High-order statistical pooling~\cite{michieli2023online}
uses temporal feature statistics for continual KWS.
Together, these studies highlight the potential of matching
and temporal aggregation for achieving favorable
accuracy--parameter trade-offs in KWS.

We therefore seek a parameter-free way to summarize
temporal matching patterns.
Mapping the temporal positions of matching responses
onto the complex unit circle enables compact statistics
of their distributions and relative timing.
This motivates CircleMatch, which combines band-wise
acoustic compression with circular aggregation of
learned prototype responses.

Our main contributions are:
\begin{itemize}
	\item We design a tiny KWS architecture combining
	band-wise acoustic compression with learned class-specific
	prototype matching, and develop four variants with
	approximately 1k--7k parameters in the 12-class setting.
	
	\item We construct circular temporal statistics that
	summarize prototype response distributions and relative
	timing without additional learned parameters.
	These statistics are invariant to a common circular shift
	of the response curves.
	
	\item We demonstrate competitive accuracy--parameter
	trade-offs on four fixed-vocabulary tasks from GSC
	and the English and Spanish Micro subsets of the
	Multilingual Spoken Words Corpus (MSWC)~\cite{mazumder2021mswc}.
	Experiments with multiple training seeds are complemented
	by ablations assessing component contributions and
	a qualitative analysis of prototype responses under
	temporal transformations.
\end{itemize}

%% file: sections/03_method.tex
\section{Method}
\label{sec:method}

\subsection{Overall framework}
\label{sec:framework}

\begin{figure*}[t]
	\centering
	\includegraphics[width=0.98\textwidth]{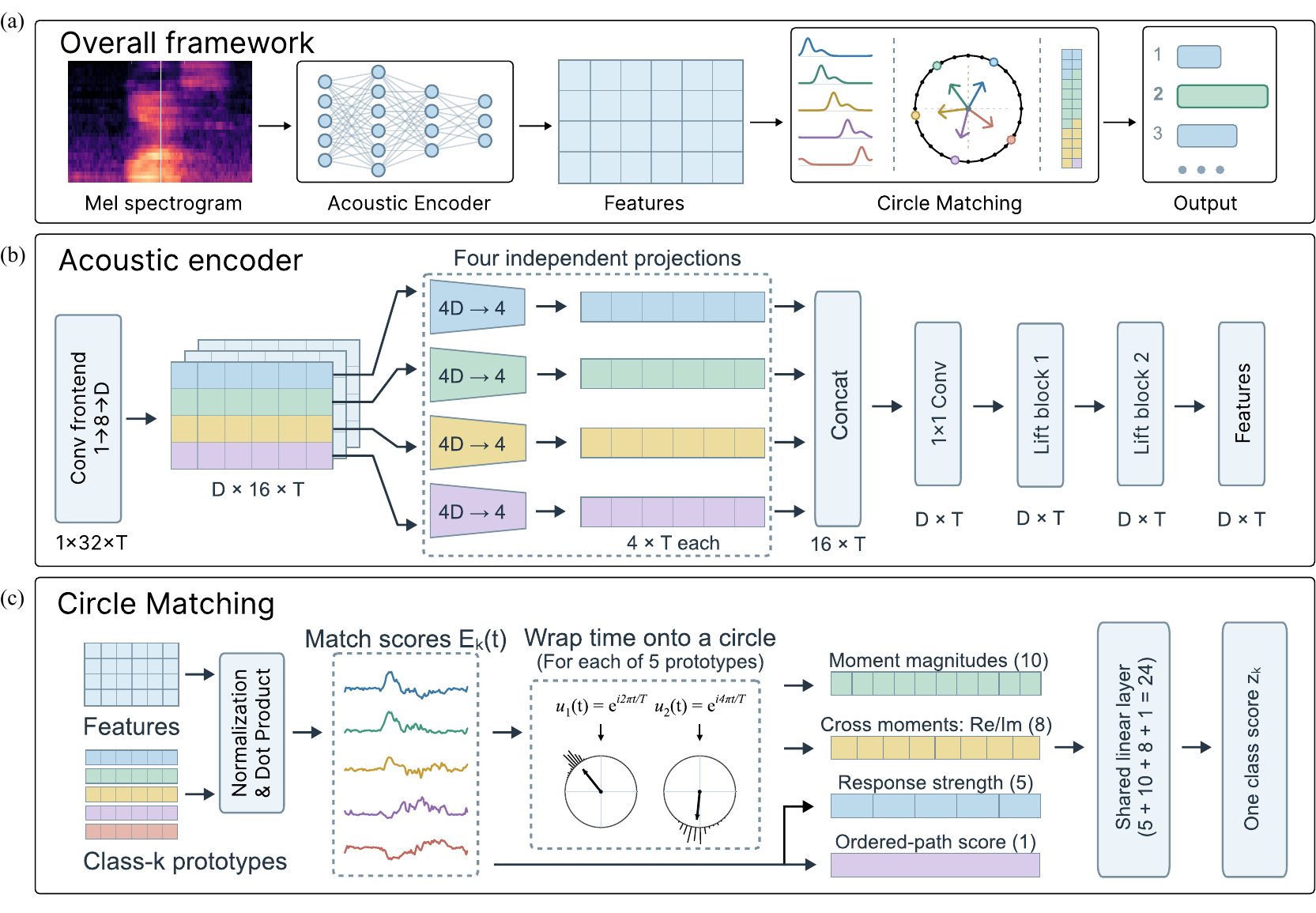}
	\par\vspace{-2pt}
	\caption{Overview of CircleMatch.
		(a) A log-Mel spectrogram is mapped to class scores through acoustic encoding and prototype matching.
		(b) Acoustic encoder with a convolutional frontend,
		four independent band projections, and two lifting blocks.
		(c) Five learned prototypes per class generate temporal responses.
		A shared linear readout combines circular statistics of responses
		normalized over time, response strength, and an ordered-path score.
		$D$ and $T$ denote feature width and frame count, respectively.}
	\label{fig:architecture}
\end{figure*}

Fig.~\ref{fig:architecture}(a) illustrates the inference
pipeline of CircleMatch.
The input audio is first preprocessed to obtain a log-Mel
spectrogram, which is then fed into the acoustic encoder
for feature extraction.
The extracted features are passed to the Circle Matching
module, where they are matched against learned acoustic
prototypes for each class.
The matching responses are mapped onto a circle according
to their temporal positions.
The resulting circular statistics are combined with
response strength and an ordered-path score to compute
a score for each class.
The highest-scoring class is selected as the prediction.

\subsection{Acoustic encoder}
\label{sec:encoder}

As shown in Fig.~\ref{fig:architecture}(b), the encoder maps
a log-Mel spectrogram $X\in\mathbb{R}^{1\times32\times T}$
to frame features $H\in\mathbb{R}^{D\times T}$,
where $T$ denotes the number of time frames and
$D$ the model width.
Its frontend uses a convolutional layer with eight output
channels, followed by a depthwise-separable convolutional
block to extract local time--frequency patterns and produce
a $D\times16\times T$ feature map.

The feature map is split into four contiguous frequency
bands, each independently compressed from $4D$ to four
dimensions per frame.
This requires $64D$ weights, versus $256D$ for a dense
$16D\!\to\!16$ projection.
The outputs are concatenated and fused by a pointwise
$16\!\to\!D$ projection.
The frontend and projection layers use batch normalization
and LeakyReLU.

Inspired by the lifting scheme~\cite{sweldens1996lifting},
we use two channel-split lifting blocks to model temporal context.
Each block splits the feature channels into
$A,B\in\mathbb{R}^{(D/2)\times T}$ and updates them as
\begin{equation}
	A'=A+\tfrac12 P(B),\qquad
	B'=B+\tfrac12 U(A').
	\label{eq:lift}
\end{equation}
After the second block, the concatenated groups form
the final encoder output frame features $H$ for prototype matching.

\subsection{Prototype matching}
\label{sec:prototypes}

Let $M=5$ denote the number of prototypes per class,
$K$ the number of classes, and $s\in\{1,\ldots,M\}$
the prototype index.
The encoder feature at frame $t$ is $h_t=H_{:,t}\in\mathbb{R}^{D}$.
For class $k\in\{1,\ldots,K\}$, $p_{k,s}\in\mathbb{R}^{D}$
is a trainable acoustic reference vector learned from
class labels without frame-level supervision.
The prototype-index order defines adjacent pairs and
the order constraint used by the ordered-path score.
As shown in Fig.~\ref{fig:architecture}(c), the matching
scores are scaled cosine similarities:
\begin{equation}
	e_{k,s,t}=\alpha\hat p_{k,s}^{\top}\hat h_t,
	\label{eq:emission}
\end{equation}
where $\hat h_t$ is obtained from $h_t$ through batch
normalization, SiLU, and $\ell_2$ normalization across
feature channels; $\hat p_{k,s}$ is the $\ell_2$-normalized
prototype; and $\alpha>0$ is a learned scale.
The components of $E_k(t)=[e_{k,1,t},\ldots,e_{k,M,t}]^\top$
form five temporal response curves per class.

Response strength is retained through five log-mean-exp
values:
\begin{equation}
	\operatorname{Strength}_{k,s}
	=\log\left(
	\frac{1}{T}\sum_{t=0}^{T-1}\exp(e_{k,s,t})
	\right).
	\label{eq:evidence}
\end{equation}

\subsection{Circular temporal statistics}
\label{sec:circular}

Beyond response strength, we summarize how responses
are distributed over time and how different prototypes
relate temporally.
We normalize each curve and use the fixed circular basis
$u_q(t)=\exp(i2\pi qt/T)$, where $q\in\{1,2\}$ and
$i^2=-1$:
\begin{align}
	w_{k,s,t}
	&=
	\frac{\exp(e_{k,s,t})}
	{\sum_{j=0}^{T-1}\exp(e_{k,s,j})},
	\label{eq:timeweights}\\
	m_{k,s}^{(q)}
	&=
	\sum_{t=0}^{T-1}w_{k,s,t}u_q(t).
	\label{eq:moment}
\end{align}
Here, $w_{k,s,t}$ is the normalized temporal weight,
and $m_{k,s}^{(q)}$ is the corresponding weighted vector
average on the circle.
The two values of $q$ wrap time once and twice,
providing complementary summaries of the response curve.

Magnitudes measure how closely the weighted circular
positions align under each mapping, while cross moments
summarize relative positions of adjacent prototype responses:
\begin{align}
	\operatorname{Magnitude}_{k,s}^{(q)}
	&=\sqrt{|m_{k,s}^{(q)}|^2+\varepsilon},
	\label{eq:magnitude}\\
	\operatorname{Cross}_{k,s}
	&=m_{k,s+1}^{(1)}\overline{m_{k,s}^{(1)}},
	\quad s=1,\ldots,M-1.
	\label{eq:cross}
\end{align}
Here, $|\cdot|$ and the overline denote complex magnitude
and conjugation, and $\varepsilon=10^{-8}$ ensures numerical
stability.
The two mappings yield ten magnitudes; four cross moments
provide eight real and imaginary components.

Under a common circular shift of the response curves,
the moment magnitudes and cross moments remain unchanged.
These 18 circular temporal statistics are computed using
a fixed basis without additional learned parameters.

\subsection{Matching evidence and classification}
\label{sec:readout}

To incorporate temporal order into the matching evidence,
we compute an ordered-path score.
Each admissible path
$\boldsymbol t=(t_1,\ldots,t_M)$ in
$\mathcal A=\{\boldsymbol t:0\leq t_1<\cdots<t_M<T\}$
selects one frame per prototype in sequence,
allowing arbitrary gaps.
We aggregate these paths as
\begin{equation}
	o_k
	=\frac{\tau_p}{M}\log\left[
	\frac{1}{\binom{T}{M}}
	\sum_{\boldsymbol t\in\mathcal A}
	\exp\left(
	\frac{\sum_{s=1}^{M}e_{k,s,t_s}}{\tau_p}
	\right)\right].
	\label{eq:path}
\end{equation}
Here, $\tau_p=0.25$ controls aggregation softness,
and $\binom{T}{M}=|\mathcal A|$ is the number of
admissible paths.
Prefix log-sum-exp dynamic programming evaluates
all class path scores in $O(KMT)$ time.
Backpropagation through this score couples prototype updates
according to their participation in ordered paths.
This guides the temporal organization of prototype responses
using only class labels, without frame-level supervision.

As shown in Fig.~\ref{fig:architecture}(c), we summarize
the matching evidence for each class in a 24-dimensional
feature vector $f_k$:
\begin{equation}
	f_k=\big[
	\underbrace{\operatorname{Strength}_k}_{5};
	\underbrace{\operatorname{Magnitude}_k}_{10};
	\underbrace{\operatorname{Cross}_k}_{8};
	\underbrace{o_k}_{1}
	\big]\in\mathbb R^{24},
	\label{eq:descriptor}
\end{equation}
The resulting feature vector is converted into a class score
as $z_k=\beta^\top f_k+b_k$, where the feature weights
$\beta\in\mathbb R^{24}$ are shared across classes and
$b_k$ is a class-specific bias.
The highest-scoring class is selected as the prediction.
All learned parameters are jointly optimized using cross-entropy.

%% file: sections/04_experiments.tex
\section{Experiments}
\label{sec:experiments}

\subsection{Experimental setup}
\label{sec:setup}
We evaluate GSC v1/v2~\cite{warden2018speech} on the official
splits with ten keywords, unknown words, and zero-waveform silence.
Each background class contains approximately 10\% as many
examples as the ten keywords combined.
MSWC~\cite{mazumder2021mswc} EN31/ES20 use the released
train/dev/test manifests and contain 31 English/20 Spanish
classes without added background classes.

Audio is resampled to 16\,kHz, cropped to the first second
or zero-padded, and converted to 32-band log-Mel features
(25\,ms window, 10\,ms hop).
Training uses $\pm100$\,ms random shifts and white-noise mixing
(probability 0.8; noise level uniform over $[-90,-46]$\,dBFS).

All models train from scratch for 100 epochs using AdamW
(batch size 256, weight decay $10^{-4}$) and cosine OneCycle
scheduling (peak learning rate $3\times10^{-3}$, 10\% warmup).
Label smoothing and dropout are 0.05 and 0.1.
Results are reported as mean ± SD over five runs on the datasets’ official test splits.

\subsection{Results and comparisons}
\label{sec:results}

Table~\ref{tab:main_results} reports accuracy and total
parameter counts for the four CircleMatch variants.
Accuracy consistently improves with model width, while
all variants remain below 10k parameters across the four tasks.
Figs.~\ref{fig:gsc_tradeoff} and~\ref{fig:mswc_tradeoff}
compare their accuracy--parameter trade-offs with published
results on GSC and MSWC, respectively.

\input{tables/main_results}
\input{figs/pareto_joint/gsc_figure}
\input{figs/pareto_joint/mswc_figure}

\subsubsection{Comparison on GSC}
\label{sec:gsc_comparison}

GSC is a competitive benchmark for tiny KWS.
As shown in Fig.~\ref{fig:gsc_tradeoff}, Circle-D4 and
Circle-D8 improve the reported Pareto frontier at the
smallest parameter budgets.
Circle-D4 uses 30.6\% fewer parameters than SparkNet-C4~\cite{svirsky2024sparse} with 3.44-point higher accuracy
on both tasks.
Circle-D8 uses 25.7\% fewer parameters than SparkNet-C8,
with comparable v2 accuracy and a 0.49-point decrease on v1.
Circle-D16 provides a modest further improvement, while
Circle-D32 remains near the frontier.
Although larger models retain higher absolute accuracy,
CircleMatch offers its strongest advantage at tiny scales.

\subsubsection{Comparison on MSWC}
\label{sec:mswc_comparison}

MSWC provides another benchmark for KWS, but published
studies use it under a broader range of training protocols
than GSC.
TRILLsson~\cite{shor2022trillsson} uses frozen pretrained
encoders with linear classifiers, while
BRILLsson~\cite{lee2024distilled} reports distilled binary
encoders and TRILL/TRILLsson baselines.
We also include full-label AST from active
learning~\cite{lee2023aloe}, the jointly trained final
exit E5 of binary early-exit networks~\cite{saeed2022binary},
and YAMNet's from-scratch closed-set EN31 result from an
audio out-of-distribution study~\cite{bukhsh2023ood}.

Although their protocols and parameter-count scopes differ,
several methods use binarization, distillation, or adaptive
exits for resource-efficient recognition.
We therefore include them in Fig.~\ref{fig:mswc_tradeoff}
as contextual comparisons.
Among the reported from-scratch results, Circle-D16 reaches
90.29\% on EN31, slightly above YAMNet's 90.05\%.
Circle-D32 further reaches 92.87\% on EN31 and 93.66\%
on ES20.
These accuracies approach or exceed those of several
much larger pretrained encoders with linear classifiers,
highlighting CircleMatch's potential for parameter-efficient
KWS using tiny complete models without pretraining or quantization.

\subsection{Responses to temporal transformations}
\label{sec:temporal_transformations}
\begin{figure}[t]
	\centering
	\includegraphics[width=\columnwidth]
	{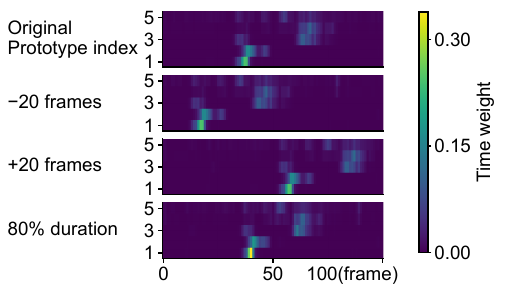}
	\par\vspace{-2pt}
	\caption{Normalized temporal weights $w_{k,s,t}$ for five
		``yes'' prototypes: original, $\pm20$-frame shifts, and
		compression to 80\% (top to bottom).
		Shifts use cropping and silence padding; scales are shared.}
	\label{fig:temporal_transformations}
\end{figure}

We apply $\pm20$-frame shifts and centered temporal compression
to 80\% of the original frame count to the log-Mel input
of one GSC v2 validation utterance of ``yes.''
Circle-D32 independently re-encodes each transformed input.
Fig.~\ref{fig:temporal_transformations} shows the normalized
temporal weights $w_{k,s,t}$; indices 1--5 identify the five
learned ``yes'' prototypes.
Peaks follow the shifts and move closer under compression
while largely retaining their order, qualitatively suggesting
shift-equivariant responses and adaptation to temporal
compression in this example.

\begin{table}[t]
	\centering
	\normalsize
	\caption{GSC v2 ablations: five-seed mean $\pm$ SD of test accuracy.
	$\Delta$: change from full D16 (percentage points).}
	\label{tab:ablation}
	\setlength{\tabcolsep}{4pt}
	\renewcommand{\arraystretch}{1.1}
	\begin{tabular}{lrrr}
		\hline
		Configuration & Params & Accuracy (\%) & $\Delta$ \\
		\hline
		Full Circle-D16
		& 3,238 & $94.99\pm0.28$ & --- \\
		w/o circular statistics
		& 3,238 & $93.72\pm0.79$ & $-1.27$ \\
		w/o cross moments
		& 3,238 & $94.56\pm0.49$ & $-0.43$ \\
		w/o ordered-path score
		& 3,238 & $94.38\pm0.35$ & $-0.62$ \\
		\hline
		Dense band projection
		& 6,310 & $95.63\pm0.37$ & $+0.64$ \\
		Mean pooling + linear
		& 2,412 & $92.06\pm0.47$ & $-2.93$ \\
		\hline
	\end{tabular}
\end{table}

\subsection{Ablation study}
\label{sec:ablation}

Table~\ref{tab:ablation} reports five-seed D16 ablations
on GSC v2. Feature masks apply during training and
inference without removing parameters.

With the same encoder, replacing Circle Matching with
mean pooling and a linear classifier reduces accuracy
by 2.93 points. Circle-D8 achieves comparable accuracy
with 29.4\% fewer parameters than this baseline
(Table~\ref{tab:main_results}).
Masking circular statistics, the ordered-path score,
and cross moments reduces accuracy by 1.27, 0.62,
and 0.43 points, respectively.
Dense projection gains 0.64 points with 95\% more parameters,
supporting efficient band-wise compression.

%% file: tables/main_results.tex
\begin{table}[t]
	\centering
	\normalsize
	\caption{Test accuracy (\%, mean $\pm$ sample SD over
		five runs) and parameter counts, including the encoder
		and classifier.}
	\label{tab:main_results}
	\setlength{\tabcolsep}{1.5pt}
	\renewcommand{\arraystretch}{1.12}
	\begin{tabular*}{\columnwidth}
		{@{\extracolsep{\fill}}lcccc@{}}
		\hline
		Dataset & D4 & D8 & D16 & D32 \\
		\hline
		GSCv1
		& $85.74\pm1.00$ & $91.11\pm0.33$
		& $94.09\pm0.15$ & $95.44\pm0.30$ \\
		GSCv2
		& $86.94\pm1.28$ & $92.11\pm0.61$
		& $94.99\pm0.28$ & $96.23\pm0.08$ \\
		\textit{Params}
		& 982 & 1,702 & 3,238 & 6,694 \\
		\hline
		EN31
		& $80.01\pm0.98$ & $87.55\pm1.03$
		& $90.29\pm0.52$ & $92.87\pm0.27$ \\
		\textit{Params}
		& 1,381 & 2,481 & 4,777 & 9,753 \\
		\hline
		ES20
		& $75.67\pm2.77$ & $86.90\pm0.45$
		& $91.70\pm0.49$ & $93.66\pm0.36$ \\
		\textit{Params}
		& 1,150 & 2,030 & 3,886 & 7,982 \\
		\hline
	\end{tabular*}
\end{table}

%% file: figs/pareto_joint/gsc_figure.tex
\begin{figure}[t]
    \centering
    \includegraphics[width=\columnwidth]{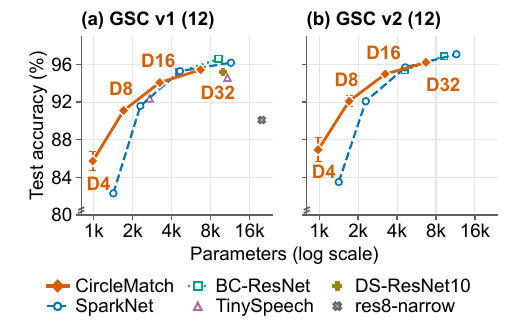}
    \par\vspace{-2pt}
    \caption{Accuracy--parameter trade-offs on (a) GSC v1 and
    (b) GSC v2, both with 12 classes.
    CircleMatch: mean $\pm$ sample SD for the runs in
    Table~\ref{tab:main_results}.}
    \label{fig:gsc_tradeoff}
\end{figure}

%% file: figs/pareto_joint/mswc_figure.tex
\begin{figure}[t]
    \centering
    \includegraphics[width=\columnwidth]{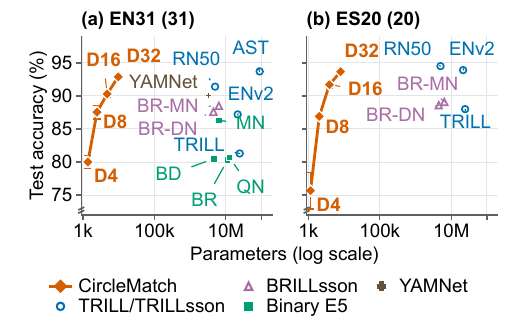}
    \caption{Accuracy--parameter trade-offs on MSWC EN31 and ES20.
    	QN, BR, BD, and MN denote QuickNet, BiRealNet,
    	BinaryDenseNet, and MeliusNet, respectively.}
    \label{fig:mswc_tradeoff}
\end{figure}

%% file: sections/05_conclusion.tex
\section{Conclusion}
\label{sec:conclusion}

We presented CircleMatch, a tiny KWS framework combining
band-wise compression, class-specific prototype matching,
and parameter-free circular aggregation.
Across four GSC and MSWC tasks, it achieves competitive
accuracy--parameter trade-offs, with its strongest advantage
at the smallest GSC budgets and promising performance across
languages and vocabulary sizes without pretraining.
Ablations support the contributions of circular statistics
and the ordered-path score, and the parameter efficiency of
band-wise compression, while the transformation example suggests
approximate shift equivariance and compression adaptation of
prototype responses.
Together, these findings suggest that fixed circular
aggregation can retain useful temporal information without
additional learned parameters, with potential applications
to sequence tasks involving relative event timing. Future work will assess on-device streaming efficiency.